\documentclass[12pt]{iopart}
\usepackage{graphicx}
\usepackage{bm} 
\newcommand{\citen}[1]{\cite{#1}}

\newcommand{\dotv}{\mbox{\boldmath\(\cdot\)}}
\newcommand{\cross}{\mbox{\boldmath\(\times\)} }

\newcommand{\sgn}{{\mathrm{sgn}\,}}
\renewcommand{\Re}{{\rm Re}}
\renewcommand{\Im}{{\rm Im}}
\newcommand{\sre}{{\rm sre}}
\newcommand{\Arg}{{\rm Arg}}

\newcommand{\kvec}{{\mathbf{k}}}

\newcommand{\rvec}{{\mathbf{r}}}
\newcommand{\vvec}{{\mathbf{v}}}
\newcommand{\xvec}{{\mathbf{x}}}
\newcommand{\yvec}{{\mathbf{y}}}
\newcommand{\zvec}{{\mathbf{z}}}

\newcommand{\khat}{\widehat{\kvec}}
\newcommand{\rhat}{\widehat{\rvec}}

\newcommand{\xhat}{\widehat{\xvec}}
\newcommand{\yhat}{\widehat{\yvec}}
\newcommand{\zhat}{\widehat{\zvec}}
\newcommand{\kDe}{k_{\rm De}}
\newcommand{\kDi}{k_{\rm Di}}

\newcommand{\ompi}{\omega_{\rm pi}}

\begin{document}

\title[Dressed test particles]{Dressed test particles, oscillation centres and pseudo-orbits}

\author{R L Dewar and D Leykam}

\address{Research School of Physics and Engineering, The Australian University, Canberra 0200, Australia}
\ead{robert.dewar@anu.edu.au}
\begin{abstract}
A general semi-analytical method for accurate and efficient numerical calculation of the dielectrically screened (``dressed'') potential around a non-relativistic test particle moving in an isotropic, collisionless, unmagnetised plasma is presented. The method requires no approximations and is illustrated using results calculated for two cases taken from the MSc thesis of the first author: test particles with velocities above and below the ion sound speed in plasmas with Maxwellian ions and warm electrons. The idea that the fluctuation spectrum of a plasma can be described as a superposition of the fields around \emph{non-interacting} dressed test particles is an expression of the quasiparticle concept, which has also been expressed in the development of the oscillation-centre and pseudo-orbit formalisms.
\end{abstract}

\submitto{\PPCF}
\maketitle

\section{Introduction}

The Fokker--Planck--Landau equation incorporating dielectric screening now known as the Balescu--Lenard equation was also derived by Thompson and Hubbard \cite{Thompson_Hubbard_60,Hubbard_61a,Hubbard_61b} from simpler statistical physics arguments. In the Thompson--Hubbard approach the diffusion coefficient was calculated from a fluctuation spectrum obtained by superimposing the dielectrically screened fields of independently moving particles, an approach called the \emph{dressed test particle picture} by Rostoker \cite{Rostoker_64}.

In this picture, unperturbed ``test particles'' replace the actual particles in the plasma, though in reality the trajectories are perturbed slightly by the fluctuations, giving rise to the linear dielectric response and velocity-space diffusion. The dressed test particle picture was developed in Fourier, $(\omega,\kvec)$, representation rather than in real space-time, $(\xvec,t)$, and thus it was difficult to visualise the actual nature of the screened potential surrounding each particle. This created interest in inverting the Fourier transform to better understand Balescu--Lenard kinetic theory.

The first chapter of the MSc thesis of the first author \cite{Dewar_67} was on relativistic plasma response functions \cite{Dewar_77}. However Chapters 2 and 3 of the thesis were restricted to the non-relativistic case and calculated the screened potential of a dressed test particle in real space, using both asymptotic approximation methods and ``exact'' calculations in which the triple Fourier integral was evaluated in a way that exploited a highly efficient numerical algorithm for calculating a special function $\eta(z)$ defined for the purpose. 

Much of the two relevant chapters of reference \citen{Dewar_67} has recently been published more-or-less verbatim \cite{Dewar_10}, but details of the numerical approach were omitted. Our main purpose in this paper is to present the mathematical analysis and numerical method used in reference \citen{Dewar_67} in the hope they may prove useful in teaching and further research.

A survey \cite{Dewar_67} of the '60s literature on  test-particle screening  is reprinted in reference \citen{Dewar_10}. The problem was revisited occasionally in the '70s, e.g. \cite{Chen_Langdon_Lieberman_73}, and '80s, but in the last decade and a half there has been a considerable revival of interest in this topic in the context of dusty plasmas. Some of these more recent papers are mentioned briefly below.

Ishihara and Vladimirov \cite{Ishihara_Vladimirov_97} applied screened potentials to charged dust particles in a plasma. They calculated the potential in the wake of a moving test particle and showed it contained periodic minima. This can result in an attractive force between dust particles, providing a mechanism for the formation of Coulomb crystals. Other studies have considered the effect of Landau damping \cite{Lampe_etal_00, Bose_Janaki_05}, neutral collisions \cite{Lampe_etal_00}, magnetic fields \cite{Nambu_Salimullah_Bingham_01} and the wake of a dipole \cite{Ishihara_Vladimirov_Cramer_00}. Lapenta \cite{Lapenta_00} considered a derivation of the screened potential in real space, rather than Fourier space, with the aim of including nonlinear effects in the wake. Numerical simulations have played an important role in validating analytic results and finding potentials beyond the point particle approximation for objects such as rods \cite{Miloch_Vladimirov_Pecseli_Trulsen_08}) and multiple particles \cite{Lampe_Joyce_Ganguli_05, Miloch_Vladimirov_Pecseli_Trulsen_09}.

While particle-in-cell simulations \cite{Winske_etal_00, Miloch_Vladimirov_Pecseli_Trulsen_08} allow more physics to be included, including nonlinearity \cite{Guio_etal_08,Hutchinson_11}, the approaches based on linear dielectric response, e.g. \cite{Chen_Langdon_Lieberman_73,Ishihara_Vladimirov_97,Bose_Janaki_05,Lampe_Joyce_Ganguli_05}, in principle allow more resolution and accuracy (within the linear r\'egime) and should be numerically less intensive. However, to simplify the problem the dielectric response function is usually approximated and/or integrations are performed only approximately.

An exception appears to be the work of Lampe and Joyce \cite{Lampe_etal_00,Lampe_Joyce_Ganguli_05}, who used Fast Fourier Transforms to perform the Fourier inversions without the need to make approximations. However, because the inversions were performed only once and tabulated to avoid the need for recalculation, calculational details and timings were not given in these papers.

In this paper we present the efficient numerical approach \cite{Dewar_67} that allowed an extensive numerical study even using the limited computer power available in the mid-'60s. The method does not require the dielectric response function to be approximated, so it includes Landau damping accurately, and was applied to dielectric constants corresponding to both Lorentzian and Maxwellian plasma distribution functions. We have found that the method allows a full, accurate calculation of the wake structure to be computed in a few minutes on a modern laptop. 

Details of the numerical techniques used and the special function $\eta$ were omitted from reference \citen{Dewar_10} so they are, until now, unpublished. In \sref{sec:dressed}  we review the Fourier representation of dielectric screening and present the strategy introduced in reference \cite{Dewar_67} for performing the 3-dimensional integral over $\kvec$ in spherical polars, by evaluating the infinite integral over $k$ first. 
This is the most difficult part of the numerical Fourier inversion as numerical integration of rapidly oscillatory integrands is notoriously difficult, but this difficulty is circumvented by drawing on the mathematical tools of complex analysis and the literature on special functions to develop an efficient algorithm for calculating this integral in terms of a special function, $\eta(z)$, thus effectively performing the $k$-integral analytically.

In \sref{sec:numerics} we introduce the coordinate system used to perform the remaining 2-dimensional integral over solid angle numerically, and in \sref{sec:results}  we indicate how this is implemented in a reconstruction of the code, using \texttt{gfortran} \cite{gfortran_11}. Some of the previous cases studied \cite{Dewar_67,Dewar_10} are recalculated and replotted as validation of the reconstructed code and a few new, but related plots are presented as well.  \ref{sec:eta} gives details of the special function $\eta(z)$ and its relation to the exponential integral $E_1(z)$ and the auxiliary function for sine and cosine integrals, $f(z)$. A subroutine for calculating $\eta(z)$ is provided online as supplementary material.

In \sref{sec:memes} we attempt to trace how the patterns of thought developed from the MSc studies of the first author have influenced the course of his career. 

\section{Dressed test particles}
\label{sec:dressed}

A test particle with charge $q$ moving through a plasma at constant, non-relativistic velocity $\mathbf{v}_0$ produces the following dielectrically screened potential \cite{Dewar_10} (in SI units) 
\begin{equation}\label{eq:response}
	\varphi = \frac{q}{\varepsilon_0}\lim_{\lambda \to +0}\int \frac{d^{3}k}{(2\pi)^{3}}
	\frac{\exp[i\mathbf{k}\cdot(\mathbf{x-v}_{0}t) - \lambda|\kvec|]}{k^{2}\epsilon(\mathbf{k}\cdot \mathbf{v}_{0},\mathbf{k})} \;,
\end{equation}
where $\varepsilon_0$ is the permittivity of free space and $\epsilon(\mathbf{\omega},\mathbf{k})$ is the frequency, $\omega$, and wavevector, $\kvec$, -dependent plasma dielectric constant. In an isotropic, collisionless, unmagnetised plasma the dielectric constant is of the form
\begin{equation}\label{eq:PhiFormDielectric}
\epsilon(\omega,\mathbf{k}) = \epsilon(\omega,|\mathbf{k}|) = 1 + \frac{\Phi(\omega / k)}{k^{2}} \;,
\end{equation}
where $\Phi(\omega/k)$ (called the \emph{polarisation function} in references \citen{Dewar_67,Dewar_10}) will be discussed more explicitly below. For now all we need to assume is that it obey the reality condition, $\Phi(-\nu) = \Phi^*(\nu)$, and the stability condition \cite{Dewar_67,Dewar_10}, $\Re\,[\Phi(\nu)]^{1/2} > 0$, for all real finite $\nu$, where $[\Phi(\nu)]^{1/2} \equiv |\Phi(\nu)|^{1/2}\exp\frac{1}{2}i\arg\Phi(\nu)$. 
In \eref{eq:response}, $\lambda$ is a regularisation parameter required to interpret the integral as $k \equiv |\kvec| \to \infty$.

Writing $\rvec = \xvec - \vvec_0 t$, we have the potential in the rest frame of the test particle
\begin{equation}\label{eq:screenedpotl}
	\varphi(\rvec) =  \frac{q}{\varepsilon_0}\lim_{\lambda \to +0}\int \frac{d^{3}k}{(2\pi)^{3}}
	\frac{\exp(i\mathbf{k}\cdot\rvec - \lambda|\kvec|)}{k^{2} + \Phi(\khat\cdot \mathbf{v}_{0})} \;,
\end{equation}
where $\khat$ is the unit vector in the direction of $\kvec$. A key insight in reference \citen{Dewar_67} was that the 3-dimensional integral in \eref{eq:screenedpotl} could be reduced to the 2-dimensional integral below by transforming to spherical polars and integrating over $k$ using the special function $\eta(\cdot)$ defined in \eref{eq:etadef}. Making use of the identity \eref{eq:etaident} differentiated twice with respect to $\alpha = \khat\dotv\rvec + i\lambda$, with $\beta = [\Phi(\khat\cdot \mathbf{v}_{0})]^{1/2} \equiv \sqrt{\Phi}$, we find
\begin{equation}\label{eq:etascreenedpotl}
	\varphi(\rvec) =  -i\frac{q}{\varepsilon_0}\lim_{\lambda \to +0} \int_{\,\khat\in S^2}\frac{d\Omega(\khat)}{(2\pi)^{3}}\,
	\sqrt{\Phi}\,\eta''\!\left((\khat\dotv\rvec+i\lambda)\sqrt{\Phi}\right) \;,
\end{equation}
where $d\Omega(\khat)$ is an element of solid angle such that $d^3k = k^2 d\Omega(\khat)$ and $S^2$ is the unit sphere. Using the identity \eref{eq:etapreven}, which gives $\eta''(z) \equiv \eta(z) -1/z$, we can also write $\varphi$ in the form of the ``bare'' Coulomb potential $\varphi_0(r)$,
\begin{equation}\label{eq:baredef}
	\varphi_0(r) \equiv \frac{q}{4\pi\varepsilon_0 r} \;,
\end{equation}
plus a correction term, which may be interpreted as the potential of the screening charge ``dressing'' the particle,
\begin{equation}\label{eq:splitscreenedpotl}
	\varphi(\rvec) =  \varphi_0(r)  -i\frac{q}{\varepsilon_0} \int_{\,\khat\in S^2}\frac{d\Omega(\khat)}{(2\pi)^{3}}\,
	\sqrt{\Phi}\,\eta\!\left(\khat\dotv\rvec\sqrt{\Phi}\right) \;.
\end{equation}

The first term  on the RHS of \eref{eq:splitscreenedpotl}, the bare potential $\varphi_0(r)$, comes from the $1/z$ term in $\eta''(z)$, which is divergent as $z \to 0$. We used the Plemelj formula to interpret $\lim_{\lambda \to +0} 1/(\khat\dotv\rvec+i\lambda)$ as ${\cal P}/\khat\dotv\rvec -i\pi\delta(\khat\dotv\rvec)$, where $\cal P$ denotes the principal part operator and $\delta(\cdot)$ is the Dirac delta function. (The integral over the principal part term vanishes, so only the delta-function term contributes to the potential.) We have set $\lambda = 0$ in the second term, the dressing potential, as, from \eref{eq:etaseries}, the singularity in $\eta\!\left(\khat\dotv\rvec\sqrt{\Phi}\right)$ at $\khat\dotv\rvec = 0$ is sufficiently weak that regularisation is not required.

The function $\Phi(\cdot)$, as derived in the standard way from the linearised Vlasov equation, is given by
\begin{equation}\label{eq:Phidef}
	\Phi\left(\frac{\omega}{ k}\right) = \sum_{s} \omega_{{\rm p}s}^{2} \int_{-\infty}^{\infty} dv \frac{g_{s}^{\prime}(v)}{\omega/ k - v } \;.
\end{equation}
Here $\omega_{{\rm p}s}$ denotes the plasma frequency, $(e_s^2
n_s/\varepsilon_0 m_s)^{1/2}$, for species $s$, with $n_s$ the
unperturbed number density, $m_s$ the mass, $e_s$ the charge, and $g_s (v)$  the one-dimensional projection of the velocity distribution function $f_s ( \mathbf{v} )$.
For a non-relativistic Maxwellian plasma,
\begin{equation}\label{eq:Phi_Max}
\Phi_{s}\left(\frac{\omega_{{\rm p}s}}{k_{{\rm D}s}} x\right) = k_{{\rm D}s}^{2} \left[ 1 - \sqrt{2}x F\left(\frac{x}{\sqrt{2}}\right) + i\sqrt{\frac{\pi}{2}} x \exp\left(-\frac{x^{2}}{2}\right) \right]
\end{equation}
where $k_{{\rm D}s} \equiv (e_s^2
n_s/\varepsilon_0 T_s)^{1/2}$ is the inverse Debye length for species $s$, $T_s$ being the temperature in energy units, and $F(x)$ the Dawson function, defined by \cite{Dawsons_Integral}
\begin{equation}\label{eq:DawsonF}
	F(\zeta) \equiv e^{-\zeta^2} \int_{0}^{\zeta} e^{t^2} dt \;.
\end{equation}
The RHS of \eref{eq:Phi_Max} can also be written as $-(k_{{\rm D}s}^2/2)Z'(x/\sqrt{2})$ where $Z(\zeta)$ is the Plasma Dispersion Function \cite{Fried_Conte_61}.

\subsection{Limitations of the method}\label{sec:Limitations}

It should be noted that the method requires that $k^2\epsilon(\kvec\dotv \vvec_0,\kvec)$ be of the form $k^2 + \Phi(\khat)$ in order for the identity \eref{eq:etaident} to be used to evaluate the infinite integral over $k$. This limits the applicability of the method to classical, collisionless, unmagnetised plasmas as explained below. The specific form we have assumed for $\Phi$ in \eref{eq:PhiFormDielectric} and \eref{eq:Phidef} is further limited to isotropic plasmas, but the method would also apply to the anisotropic case, complicating only the integration over solid angle. 

The classical polarisation function $\Phi$  involves numerators, $\kvec\dotv(\partial f_s/\partial\vvec)$, and denominators, $\omega - \kvec\dotv\vvec$, arising from solving the linearised Vlasov equation, while in the quantum case \cite{Else_Kompaneets_Vladimirov_10} the numerators are replaced by $(m_s/\hbar)[f_ s(\vvec + \hbar\kvec/2m_s) ) - f_ s(\vvec - \hbar\kvec/2m_s)]$, which reduces to the classical form as $\hbar \to 0$. To obtain the form in \eref{eq:Phidef}, or its anisotropic generalisation, we divide both numerators and denominators by $k$, so that all $k$-dependences of the numerators are removed and the numerators depend on $k$ only through the phase velocity $\omega/k$. The problem encountered in the quantum case is that the numerators are \emph{not} linear in $\kvec$, so that their $k$-dependences are not removed by this division. The problem encountered in the collisional case of \cite{Lampe_etal_00,Lampe_Joyce_Ganguli_05}, where $\omega$ in the denominator of the ion term is replaced by $\omega + \nu_i$, where $\nu_{\rm i}$ is the ion-neutral collision frequency, is that $\nu_{\rm i}/k$ is not independent of $k$ (unless $\nu_{\rm i}$ is assumed proportional to $k$) so that the denominator no longer depends only on $\omega/k$. Similarly, in the magnetised case the cyclotron frequencies, $\omega_{{\rm c}s}$, would complicate the $k$-dependences of the denominators.

\begin{figure}[tbp]		
	\centering
	\includegraphics[width=5cm]{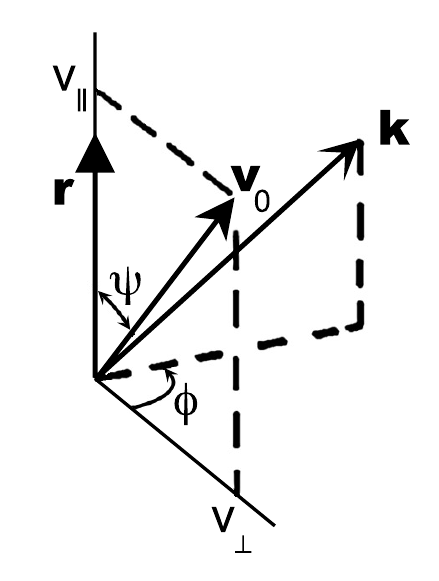}
	\caption{Coordinate system used for numerical evaluation of the solid-angle integral in \eref{eq:splitscreenedpotl}.}
	\label{fig:MScFig3_1}
\end{figure}

\section{Numerical formulation}
\label{sec:numerics}

Two systems of coordinates for evaluating the solid-angle integral in \eref{eq:etascreenedpotl} numerically suggest themselves: spherical polars with axis along $\vvec_0$, and spherical polars with axis along $\rvec$ as shown in \fref{fig:MScFig3_1}. We adopt the second choice because it appears to have the advantage that it is the natural choice when  deriving the bare potential term in \eref{eq:splitscreenedpotl}, where the singularity at $\khat\dotv\rvec = 0$ is of particular concern and $\vvec_0$ does not appear. In this coordinate system we have
\begin{equation}
	\khat\dotv\vvec_0 = \mu v_{\parallel} + \sqrt{1-\mu^2}\cos\phi\,v_{\perp} \;,
	\label{eq:kdotv}
\end{equation}
where $v_{\parallel} \equiv \vvec_0\dotv\rhat = v_0\cos\psi$, $v_{\perp} \equiv |\vvec_0\cross\rhat| = v_0\sin\psi$, and $\mu \equiv \khat\dotv\rhat$. 

\begin{figure}[tbp]
	\centering
\begin{tabular}{cc}
	\begin{tabular}{c}
		\includegraphics[width=7.5cm]{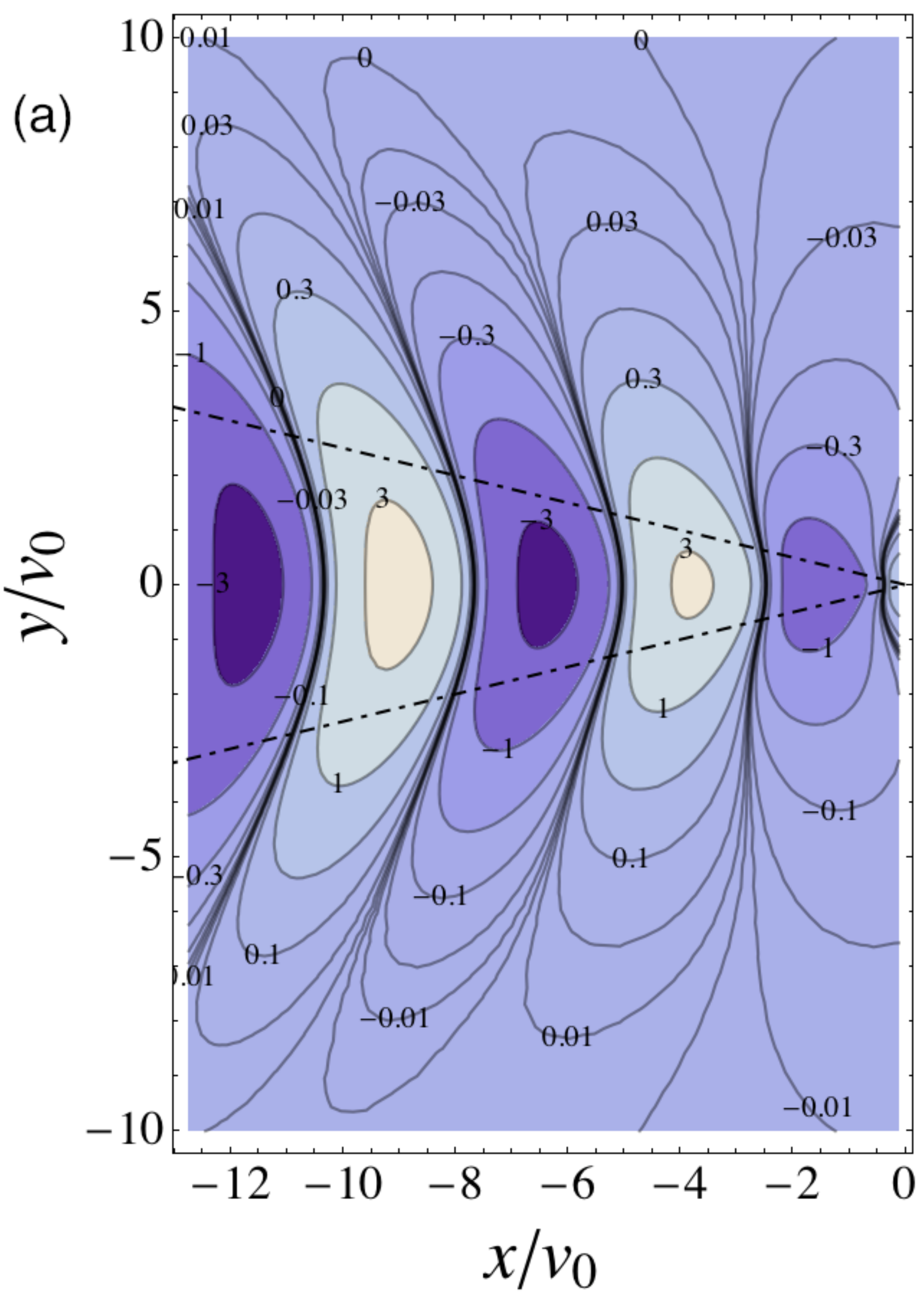}
	\end{tabular}	
&	\begin{tabular}{c}
		\includegraphics[width=6.5cm]{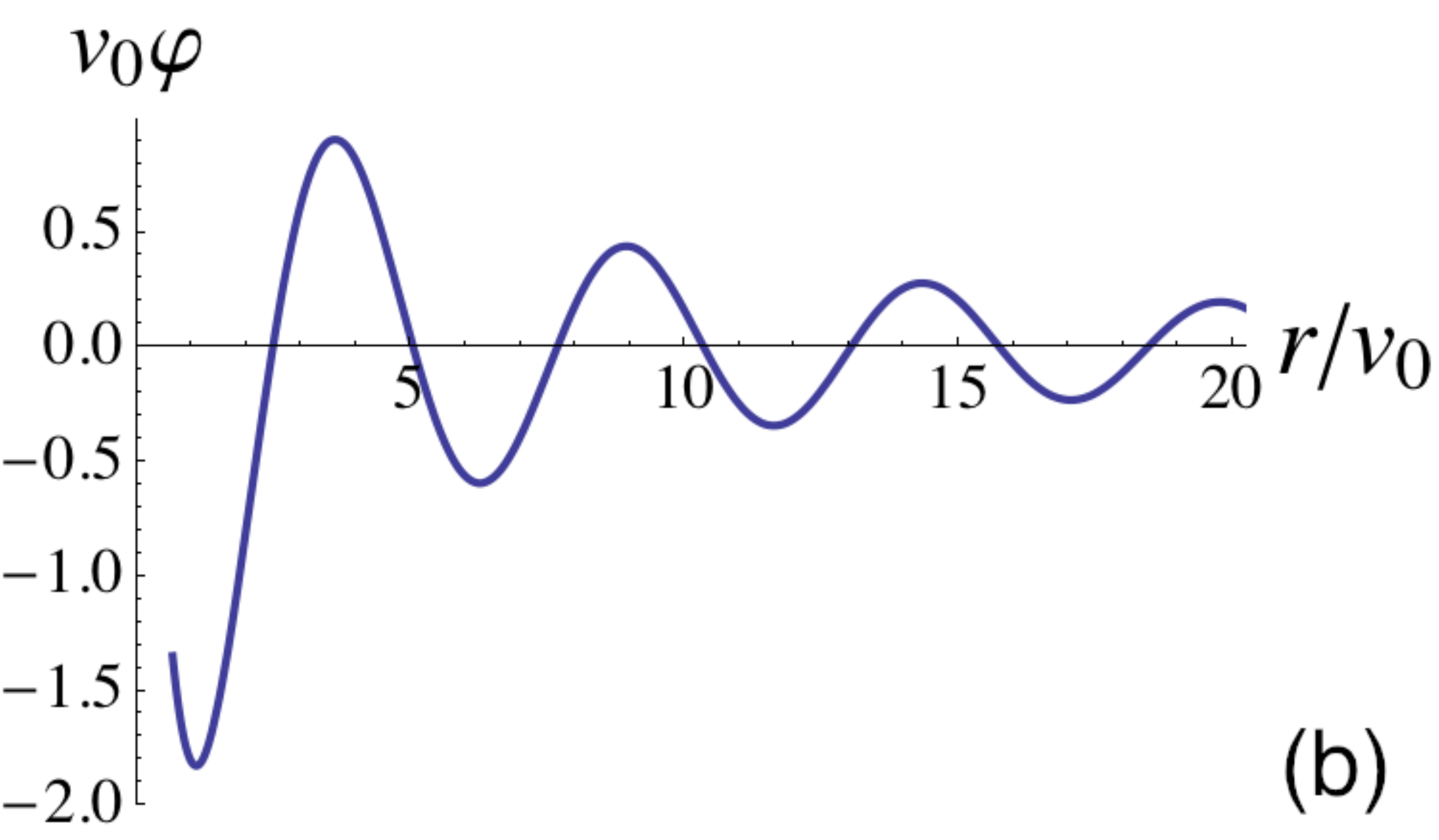}\\
		\mbox{}\\
		\includegraphics[width=6.5cm]{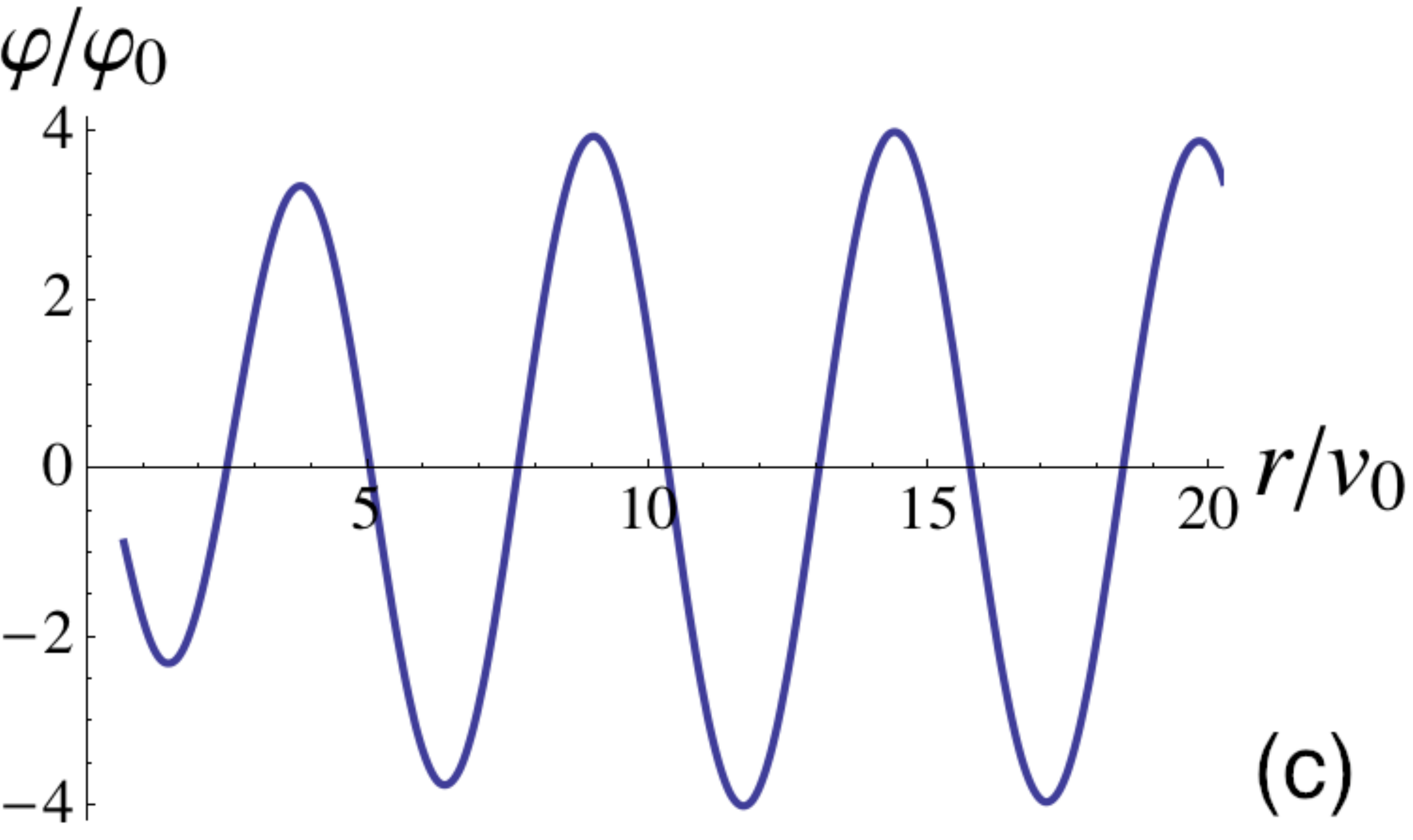}
	\end{tabular}	
\end{tabular}	
\caption{Visualisations of the wake of a superthermal but subsonic test particle moving at $v_0 = 4v_{\rm thi} \equiv 4\ompi/\kDi$ in a Maxwellian-ion plasma with $\kDe = 0$ (colour online). (a) Contour plot of $\varphi(\rvec)/\varphi_0(r)$. This figure agrees with figure~13 of \cite{Dewar_10}, the dot-dashed lines delineating the ``thermal Mach cone.'' (b) $v_0\varphi(r,\pi)$ \emph{vs.} $r/v_0$ behind the particle. This figure agrees with figure~12 of \cite{Dewar_10}. (c) $\varphi(r,\pi)/\varphi_0(r)$ \emph{vs.} $r/v_0$ behind the particle. This is a transect of (a) along the negative $y = 0$ axis, extended to $x/v_0 = -20$. }
	\label{fig:Figs12and13Dewar10}
\end{figure}

In these coordinates the element of solid angle in \eref{eq:etascreenedpotl} is given by $d\Omega = d\mu d\phi$, with the ranges of integration being $\mu \in [-1,1]$,  $\phi \in [0,2\pi]$. By using the reality condition, the ranges of the $\mu$ and $\phi$ integrations can be reduced by half, so \eref{eq:splitscreenedpotl} becomes 
\begin{equation}\label{eq:phiNum2}
	\varphi(r,\psi) = \frac{2\varphi_0(r)}{\pi^2} \int_{0}^{\pi} d\phi \left\{ \frac{\pi}{2}
	+ \int_{0}^{r} dx\, \mathrm{Im} \left[\sqrt{\Phi}\, \eta \left(x\sqrt{\Phi}\right) \right] \right\}
\end{equation}
with
\begin{equation}\label{eq:sqrtPhidef}
	\sqrt{\Phi} \equiv
	\left\{
		\Phi\left(v_0\cos\psi\,\frac{x}{r}
		+ v_0\sin\psi\cos\phi\left[1 - \left(\frac{x}{r}\right)^2\right]^{1/2}\right)
	\right\}^{1/2} \;.
\end{equation}
The first term in the integrand of the $\phi$-integral (which arises from the delta-function in the Plemelj formula) gives the bare Coulomb potential $\varphi_0$. The second term, the integral over $x \equiv r\mu$, gives the dressing potential from the screening cloud. As will be commented on further below, these two terms almost cancel in the far-field of Debye-screened regions.

\begin{figure}[!ht]
	\centering
	\begin{tabular}{cc}
	\includegraphics[width=7cm]{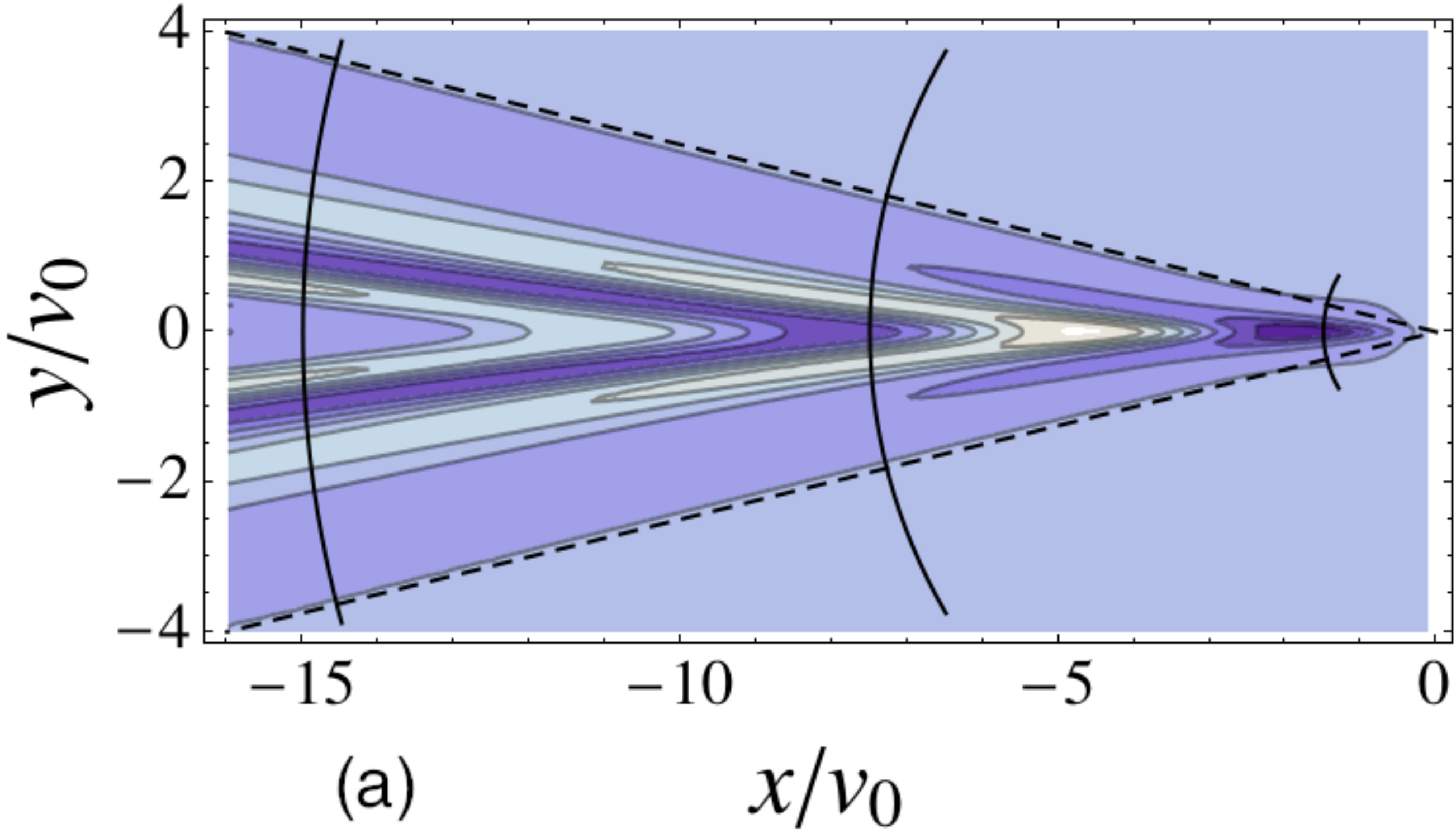}   &  \includegraphics[width=6.5cm]{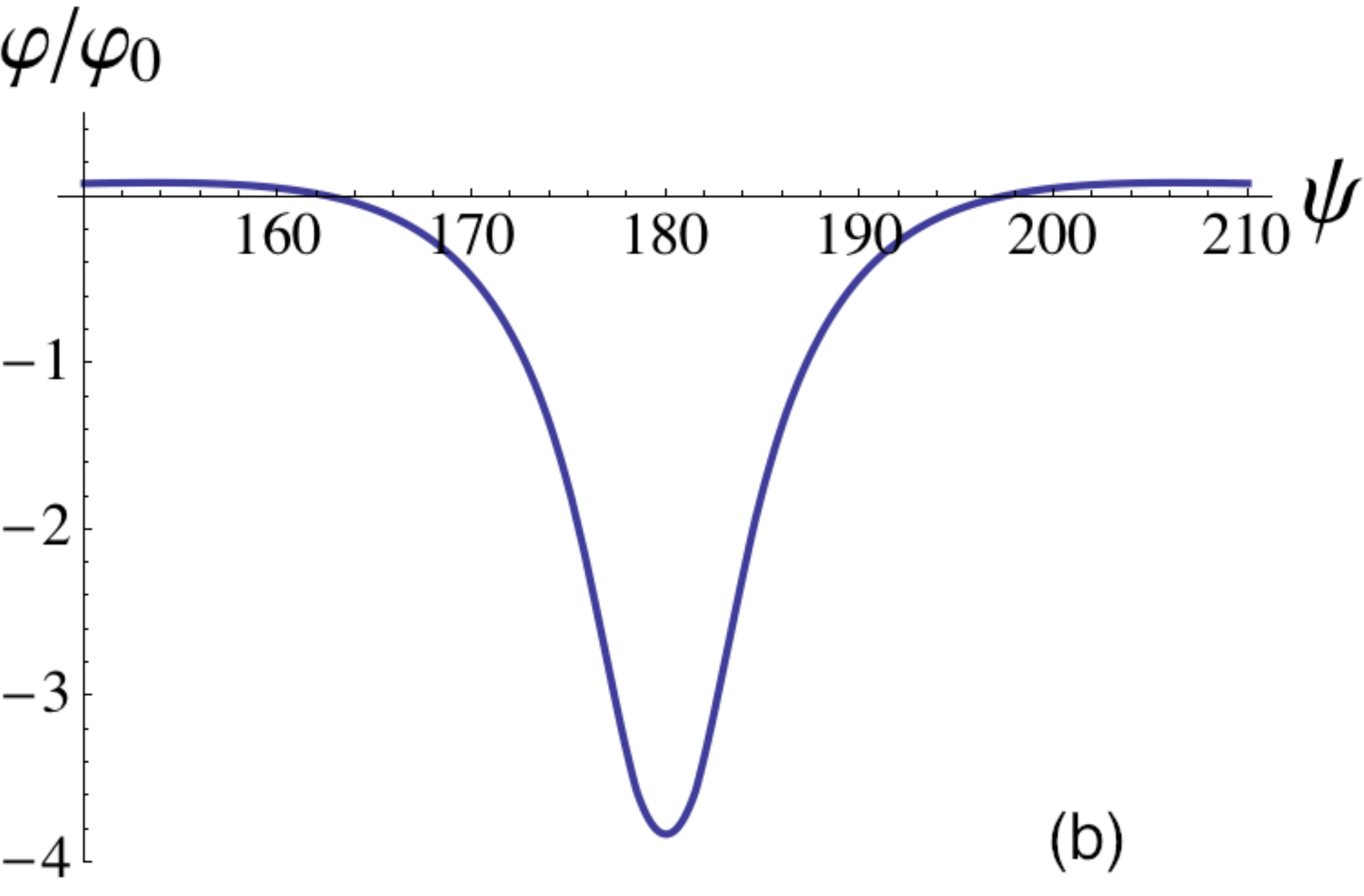}\\
	\includegraphics[width=6.5cm]{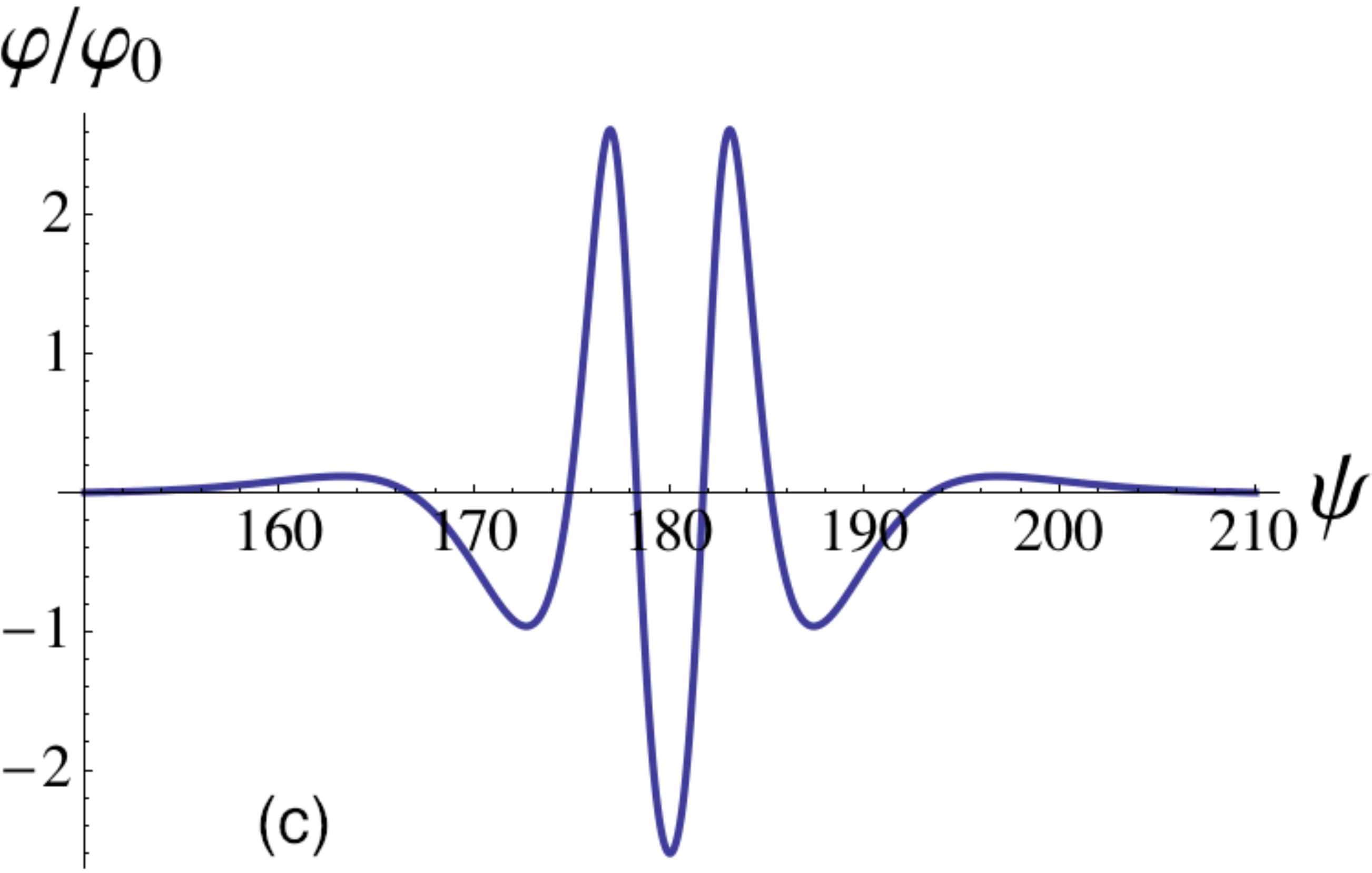}   &  \includegraphics[width=6.5cm]{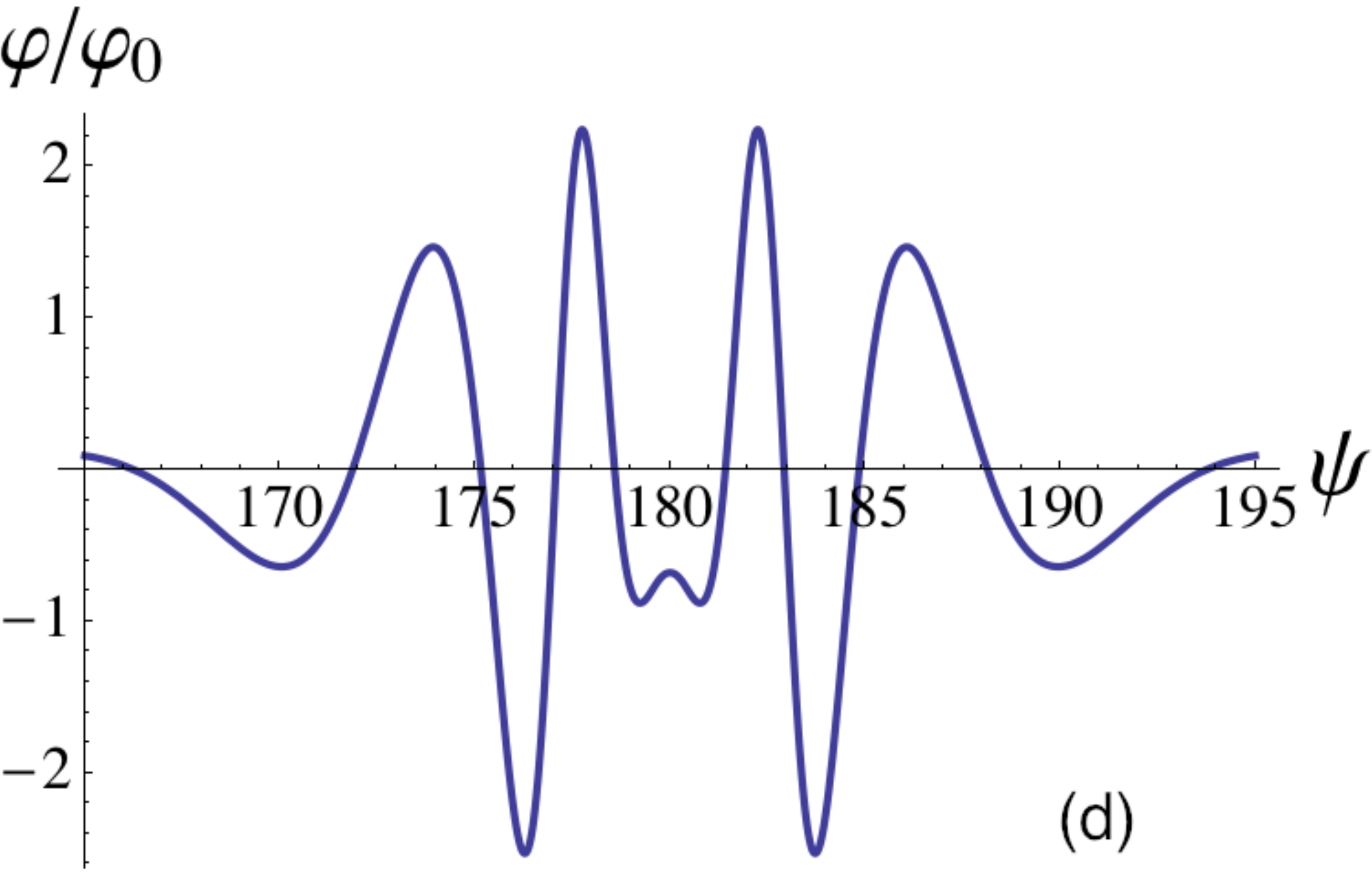}
	\end{tabular}
	\caption{Visualisations of the wake of a supersonic test particle moving at $v_0 = 32\ompi/\kDi = 4C_{\rm s}$ in a Maxwellian-ion plasma with $\kDe = \kDi/8$. (a) Contour plot of $\varphi(\rvec)/\varphi_0(r)$. (b) $\varphi(r,\psi)/\varphi_0(r)$ \emph{vs.} $\psi$ (in degrees) on the small arc $r = 1.5 v_0/\kDi$) on the right of (a). This figure agrees with figure~14 of \cite{Dewar_10}. (c) $\varphi/\varphi_0$ on the middle arc, $r = 7.5 v_0/\kDi$. (d) $\varphi/\varphi_0$ on the left-most arc, $r = 15 v_0/\kDi$.}
	\label{fig:Fig14Dewar10}
\end{figure}

\section{Numerical results:  validation against MSc plots}
\label{sec:results}

We have reused the subroutine \texttt{ETA} listed in reference \cite{Dewar_67} (a listing with a README file and a test program being provided online as supplementary material for this paper) for calculating the special function $\eta(z)$, \eref{eq:etadef}, in a FORTRAN 77 reconstruction of the original FORTRAN program for computing $\varphi(r,\psi)$ from \eref{eq:phiNum2}. This code computes the double integral over $\phi$ and $x$ using Romberg integration \cite{Press_etal_86} with the maximum size of the approximation matrix set to $10\times 10$ in both integrations. The Free Software Foundation's \texttt{gcc}-based \texttt{gfortran} \cite{gfortran_11} was used to compile and link the program.

In the calculations we used units used such that $\kDi = \ompi = 1$ and $q = 4\pi\varepsilon_0$ so that $\varphi_0(r) = 1/r$. Assuming the particle velocity to be much less than the mean electron velocity, we approximated the electron polarisation function $\Phi_{e}$ with its static value, $\Phi_{e}(0) = k_{{\rm De}}^{2}$ and used the \texttt{dawson} routine \cite{Press_etal_86} to calculate the ion polarisation function from \eref{eq:Phi_Max}.

Figures \ref{fig:Figs12and13Dewar10}(a) and \ref{fig:Fig14Dewar10}(a) show contour plots of $\varphi(\rvec)/\varphi_0(r)$ in the $x/v_0,y/v_0$ plane, where $x$ and $y$ are such that $\rvec = x\xhat + y\yhat + z\zhat$, with the unit vector $\xhat$ in the direction of $\vvec_0$. The other panels show the behaviour of $\varphi$ along transects as described in the captions. The figures are both for test particles moving faster than the ion thermal speed, \fref{fig:Figs12and13Dewar10} showing a subsonic case and \fref{fig:Fig14Dewar10} showing a supersonic case (see captions).

These two figures were produced by interpolation from $100\times 100$  grids above the $x$-axis, with the contours below the axis being obtained by reflection. The 10,000 evaluations each of the Fourier inversions \eref{eq:screenedpotl} were calculated as described above. Using \texttt{gfortran} on a MacBook Pro, the scan for \fref{fig:Figs12and13Dewar10}  took 311 seconds, while that for \fref{fig:Fig14Dewar10} took 942 seconds, giving an average of 0.03 s and 0.09 s per Fourier inversion, respectively. The interpolations and plots were done with \emph{Mathematica} \cite{Mathematica8}. 
	
In \fref{fig:Fig14Dewar10}(a) the disturbance behind the particle is seen to be limited to a Mach cone (dashed lines) of half width about 15 degrees. Note that the Mach cone does not contain a shock wave owing to the dispersive character of ion acoustic waves. 
Asymptotic analyses performed in reference \citen{Dewar_67} were published in \cite{Dewar_10}. Further analyses, including an interpretation of the wave structure observed in \fref{fig:Fig14Dewar10} based on ray tracing, will be published elsewhere. Here we limit ourselves to a qualitative heuristic interpretation of the wake.

In the rest frame of the plasma, the screening response excited by the bare potential of a test particle as it passes a given point occurs with a time lag, producing a region of opposite potential immediately behind the particle but travelling with it, as shown in \fref{fig:Fig14Dewar10}(b). In \fref{fig:Fig14Dewar10}(c,d) it is seen that the sign of this near-field disturbance is preserved away from the $x$ axis in the potential minimum just inside the Mach cone in the far field. This is a wave train consisting of a superposition of ion acoustic waves with phase velocities given by $\khat\khat\dotv\vvec_0$, diminishing in amplitude away from the test particle due both to spreading and Landau damping.

In the Debye-screened regions outside the Mach cone there is strong cancellation between the bare potential and screening cloud terms in \eref{eq:phiNum2}. Thus, when $\varphi$ is very small the relative error can become significant, leading to some numerical ``noise'' causing spurious zero contours in the upper and lower right-hand corners of \fref{fig:Fig14Dewar10}.  This was filtered out by setting $\varphi/\varphi_0$ to $0.01$ when it fell below $0.01$ in absolute value outside the Mach cone, but it may be possible to eliminate the problem by using \eref{eq:etascreenedpotl} instead of \eref{eq:splitscreenedpotl} and deforming the contour of the numerical $\mu$-integration away from origin to avoid the singularity there, thus avoiding the split between bare and screening potentials and handling Debye cancellations analytically.

\section{Quasiparticles}
\label{sec:memes}

As indicated in the Introduction, the original motivation of this work \cite[Chapters 2 and 3]{Dewar_67} was mainly to achieve a better understanding of plasma kinetic theory via a visualisation of dressed test particles. 

Another motivation was to test the calculation by Pines and Bohm \cite{Pines_Bohm_52} of dynamical screening using a classical version of their quantum-mechanical collective coordinate approach \cite{Bohm_Pines_51,Bohm_Pines_53}. Their calculation in fact failed this test \cite[p. 9]{Dewar_10}, but more recently a corrected quantum-mechanical collective coordinate approach has been developed to calculate the interaction of dust particles in plasmas \cite{Ishihara_Vladimirov_98}.

While Chapter 4 (unpublished) of \cite{Dewar_67} was on a quantum field theory treatment of electron-photon scattering in a plasma, including a test particle calculation, the first author has not used quantum mechanics \emph{per se} since, except for a short project on statistical mechanics of a thin film \cite{Dewar_Frankel_68} after submitting his MSc thesis and before commencing PhD studies at Princeton.

However his experience with the quantum approach for calculating nonlinear wave-particle and wave-wave interactions led naturally to the thought that perhaps the relative ease with which this is done in quantum field theory \cite[e.g.]{Tsytovich_77,Melrose_80} is due to the power of the Lagrangian and Hamiltonian \emph{formalisms} developed in the field, rather than being due to anything intrinsic to quantum \emph{physics}. Perhaps, if similar effort were applied to developing Lagrangian and Hamiltonian methods for classical plasmas, similarly powerful results would follow. This thought, combined with the one related to the dressed test particle picture described below, has been expressed in much of his later work. 

In  quantum field theory, especially as applied in condensed matter physics, dressed test particles are regarded as ``quasiparticles'', related to bare particles but with properties altered by the interactions between the bare particles.  In quantum theory the first step is often to write down a Hamiltonian operator consisting of an unperturbed (bare) part and an interaction term, and then to diagonalise the Hamiltonian using a Bogolyubov transformation. The analogue of this in classical Hamiltonian theory would be to find a canonical transformation to a normal form in which explicit many-body interactions were reduced to an irreducible residual part, with the collective interactions being incorporated in a renormalised ``unperturbed'' Hamiltonian.

The Pines--Bohm \cite{Pines_Bohm_52} development of collective coordinate theory for particles interacting via Coulomb potentials was very much based on classical canonical transformation theory. However, the first author began to develop a quasiparticle formalism not in a many-body Hamiltonian context, but in a Lagrangian approach to wave-background interaction in an ideal magnetohydrodynamic (MHD) fluid \cite{Dewar_70}, which showed that the conservation of wave action arises as naturally in classical mechanics as in quantum mechanics. However, another expression of the quasiparticle concept was in developing an oscillation-centre theory of turbulent phase-space diffusion \cite{Dewar_72a,Dewar_73b,Dewar_76b}, which used canonical Hamiltonian perturbation theory. The operator method for classical canonical transformations introduced in \cite{Dewar_76b} triggered the adoption of Lie methods as an important tool in theoretical plasma physics \cite{Cary_81}.

As it became apparent that the separation of the transformed Hamiltonian into a renormalised, non-interacting part and a residual interaction part was highly non-trivial and intimately related to developments in Kolmogorov-Arnold-Moser (KAM) theory, 
more recent development of this line of thought has
been in papers aimed at developing a canonical theory of almost-invariant tori in $1\frac{1}{2}$ Hamiltonian systems via a pseudo-orbit approach \cite{Dewar_85,Dewar_Meiss_92,Dewar_Khorev_95,Dewar_Hudson_Gibson_11}, with physical application to the description of magnetic fields in non-axisymmetric toroidal plasma confinement systems \cite{Dewar_Hudson_Price_94,Hudson_Dewar_96,Hudson_Dewar_98,Hudson_Dewar_99,Hudson_Dewar_09}. Whether this approach will ever succeed in finding a classical Hamiltonian derivation of the Thompson--Hubbard dressed test particle picture is yet to be seen, but the development of modern Poisson-bracket perturbation formalisms \cite{Morrison_05} may point the way. Also, recent developments in the collective coordinate approach \cite{Ishihara_Vladimirov_98} look very promising.

Other concepts picked up during the first author's MSc research, such as the use of special functions, asymptotic expansions and numerical analysis also of course have found expression in various ways in his subsequent publications but we do not attempt to analyse these here.

\section{Conclusion}
\label{sec:concl}
We have presented analytical and numerical details of a novel approach, introduced in reference \cite{Dewar_67}, to inverting the Fourier transform of the screened field of a test particle, and have illustrated the method with results from two cases, subsonic and supersonic test particles.

The lack of computational detail in the literature, and the rapid increase in computer speed over the years, makes direct comparison with other approaches difficult. However, the approach presented here, being based on classical analysis techniques for special functions and sound numerical analysis principles, is expected to be relatively fast. Anecdotal evidence appears to bear this out. The reason that the method has not been presented before is presumably that the ``auxiliary function for sine and cosine integrals,'' $f(z)$ \cite{Abramowitz_Stegun_72}, which could have been used instead of our function $\eta(z)$, is not well known in the physics community, and no subroutines for its rapid evaluation have previously been available. A primary purpose of this paper is to increase knowledge of this function and perhaps stimulate the development of better subroutines than the one we have provided online (see the online README file for error plots).

The use of this special function imposes certain restrictions on the applicability of the method, making it inapplicable to quantum, magnetised and collisional plasmas as discussed in \sref{sec:Limitations}.

In \sref{sec:memes} we have also sketched how the dressed-test-particle picture is related to the concepts of quasiparticles, oscillation centres and pseudo-orbits. This section has been included to help relate the present paper to others in the cluster of papers for which it is intended.

\ack
The first author acknowledges the inspiration and encouragement of his thesis supervisors, K.~C. Hines and R.~M. Kulsrud, and from his fellow researchers and research students over the years. In particular, he wishes to acknowledge L. Chen, I.~H. Hutchinson, G. Joyce, M. Lampe, A.~B. Langdon, M.~A. Lieberman and S.~V. Vladimirov for information about relevant publications.

\begin{figure}[!ht]
	\centering
\begin{tabular}{cc}
	\includegraphics[width=6.5cm]{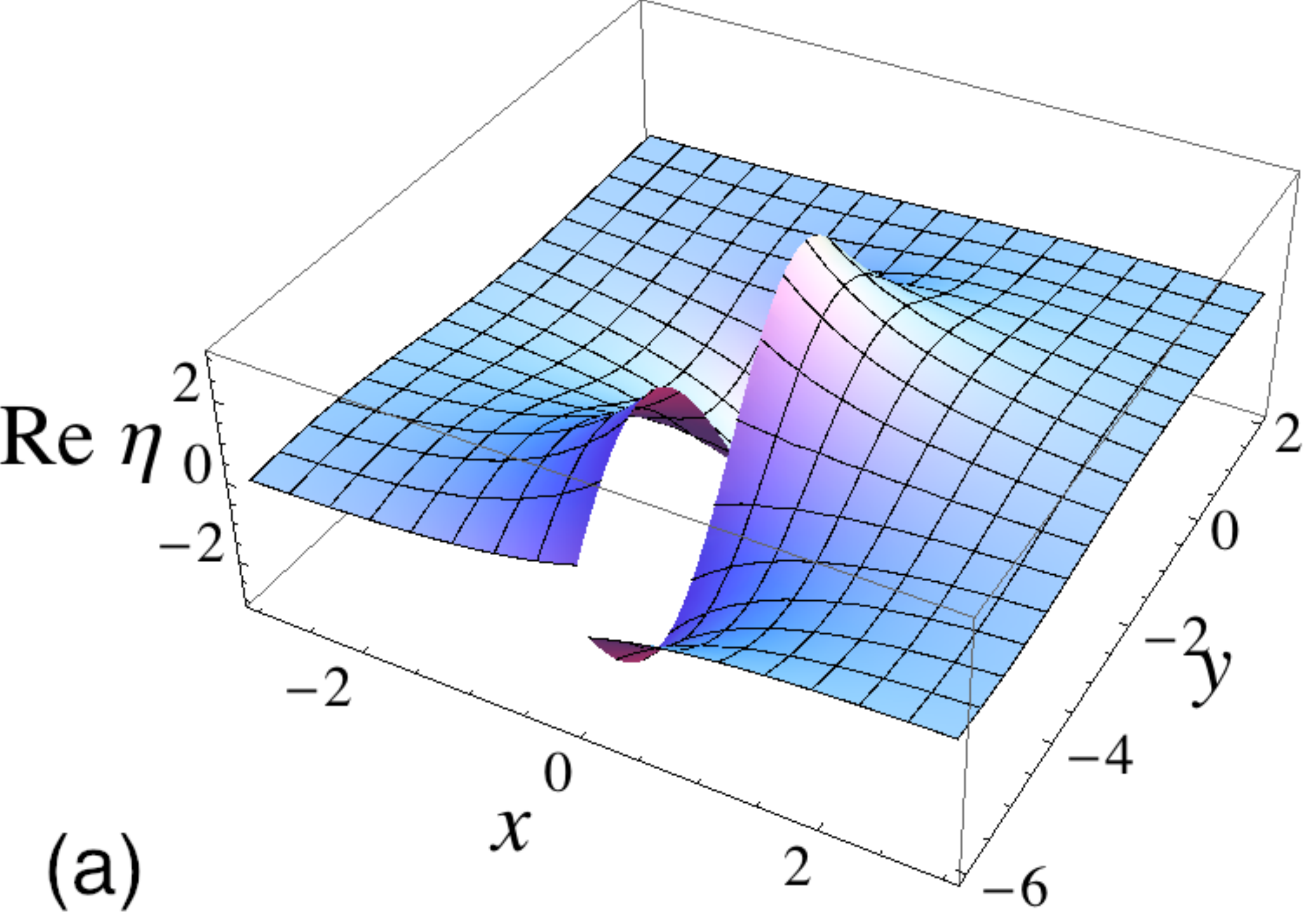}
&	\includegraphics[width=6.5cm]{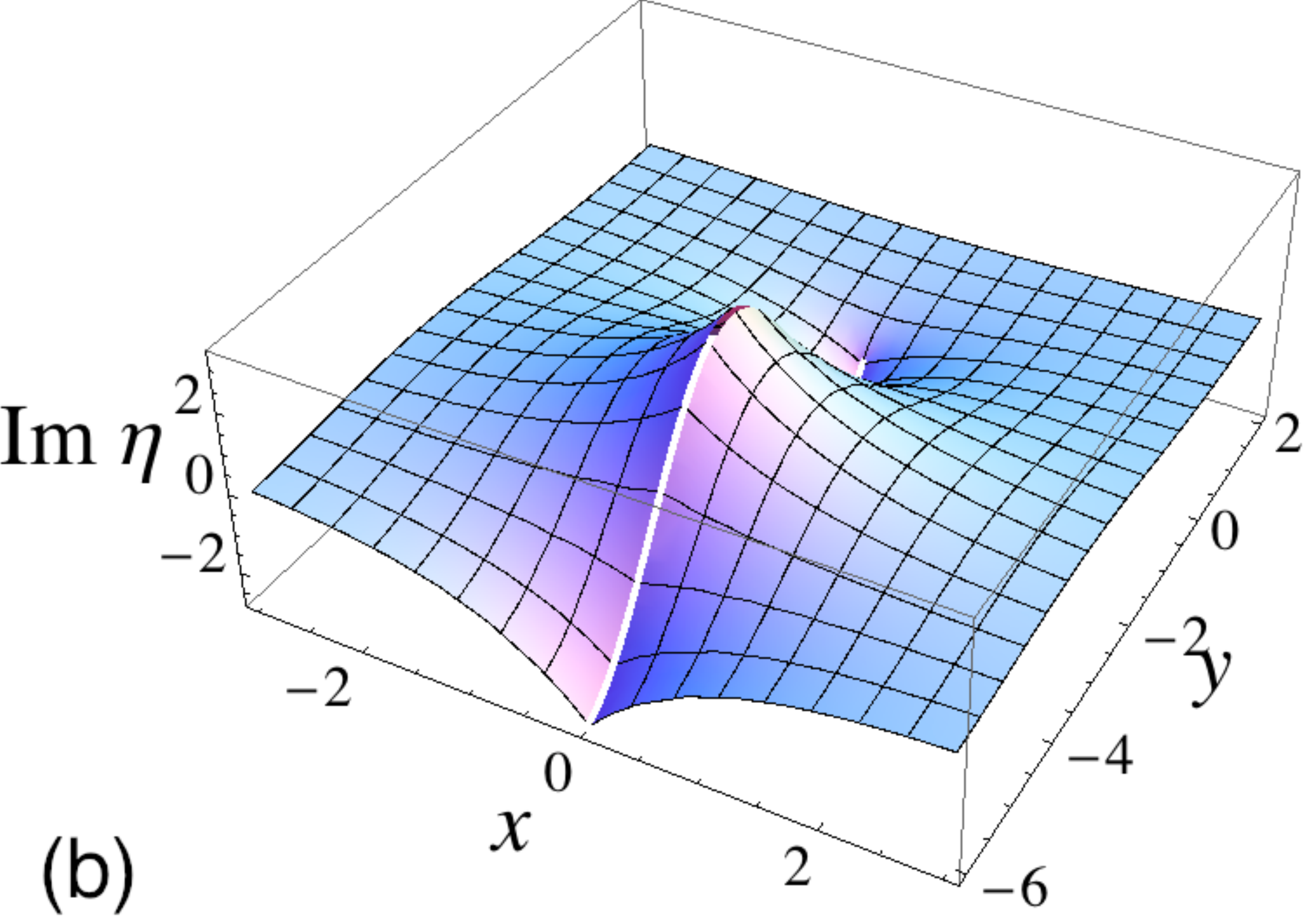}\\
\end{tabular}	
\caption{(a) Real and (b) imaginary parts of $\eta(x+iy)$ showing their respective antisymmetry/symmetry \eref{eq:ImAxisRefl} under reflection in the imaginary axis of the $(x+iy)$-plane, cut along the negative imaginary axis.}
	\label{fig:etaplots}
\end{figure}

\appendix
\section{The function $\eta$}
\label{sec:eta}

The function $\eta (z)$, first introduced in reference \citen{Dewar_67}, was used in both the analytical and numerical work in the unpublished thesis. Much of reference \citen{Dewar_67} was published in reference \citen{Dewar_10}, but Appendix~I, giving details of the mathematical properties and numerical calculation of $\eta (z)$, was omitted. Thus we reproduce the contents of Appendix~I below, including a small amount of additional material for clarity.

The function is closely related to the exponential integral \cite[\S 5.1.1]{Abramowitz_Stegun_72}, and may be computed from tables of this function. As shown below it is even more closely related to the auxiliary function for sine and cosine integrals, $f(z)$ \cite[\S 5.2.6]{Abramowitz_Stegun_72} and \cite{AuxiliaryFn_f}, but does not appear to have been defined before so the notation is our own.

Define $\eta (z)$ by
\begin{equation}\label{eq:etadef}
	\eta (z) = \frac{z}{i}\int_{0}^{\infty } dx\, \frac{e^{ix}}{x^{2} + z^{2}}
	\quad\mathrm{for}\quad |\arg z| <\frac{\pi}{2} \;,
\end{equation}
and analytically continue the function into the left half plane, cutting the complex plane along the negative imaginary axis (see \fref{fig:etaplots}).

It may be shown that
$\eta(z)$ has the alternative integral representation
\begin{equation}\label{eq:alt}
	\eta (z) = z \int_{0}^{\infty } dt\, \frac{e^{-t}}{z^{2} - t^{2}}
	\quad\mathrm{for}\quad 0 < \mathrm{arg} z< \pi \;.
\end{equation}

The form most useful for our purposes is the identity
\begin{equation}\label{eq:etaident}
	\frac{\beta}{i}\int_{0}^{\infty } \frac{\exp i\alpha x}{x^{2} + \beta^{2}} \,dx
	\equiv \eta(\alpha\beta)
	\quad\mathrm{for}\quad \Im\,\alpha \geq 0, \quad  \Re\,\beta > 0 \;.
\end{equation}
For instance, we can relate $\eta(\cdot)$ to the auxiliary function $f(z)$ \cite[\S 5.2.12]{Abramowitz_Stegun_72}
by putting $\beta = 1$, $\alpha = iz$ to give $f(z) = i\eta(iz)$ for $\Re\, z > 0$. Then, as $f(z)$ is defined on the complex $z$-plane cut along the negative \emph{real} axis \cite[\S\S 5.2.2, 5.2.6]{Abramowitz_Stegun_72}, replacing $z$ with $-iz$ we have, by analytic continuation, an alternative definition for $\eta(z)$,
\begin{equation}\label{eq:fident} 
	\eta(z) = -if(-iz)  \quad\mathrm{for}\quad -\frac{\pi}{2} < \Arg\,z < \frac{3\pi}{2} \;,
\end{equation}
the cut now being along the negative \emph{imaginary} axis as stated after \eref{eq:etadef}.

As will be found useful in the subroutine to be described below, we may also show from \eref{eq:alt} that
\begin{equation}\label{eq:expint}
	\eta ( z ) = \frac{1}{2} [ e^z E_{1}^{(+)}(z) - e^{-z} E_{1}^{(-)}(-z)] \;,
\end{equation}
where $E_{1}^{(\pm)}(z)$ are analytic continuations of the exponential integral $E_1(z)$ \cite[\S 5.1.1]{Abramowitz_Stegun_72} and \cite{DLMF_Exponential_Integrals} to the $z$-planes cut along the negative/positive imaginary axis (analytic in the upper/lower half planes), respectively [see below \eref{eq:E1series}].

These functions can be represented as, 
\begin{eqnarray}\label{eq:E1series}
	E_{1}^{(\pm)}(z) & = & -\gamma - \ln^{(\pm)} ( z ) - \sum_{n=1}^{\infty} \frac{(-z)^n}{n n!} \nonumber \\
			  & \equiv & -\gamma - \ln^{(\pm)} ( z ) + \mathrm{Ein}(z)\;
\end{eqnarray}
where $\gamma$ is Euler's constant, Ein$(z)$ \cite{DLMF_Exponential_Integrals} is an entire function, and $\ln^{(\pm)}(z) \equiv \ln|z| + i[\arg(\mp iz) \pm\pi/2] = \ln z  + i[\arg(\mp iz) - \arg z \pm\pi/2] $. Thus, comparing \eref{eq:E1series} with reference \cite[\S 5.1.1]{Abramowitz_Stegun_72},
\begin{equation}\label{eq:E1contin}
	E_{1}^{(\pm)}(z) = E_{1}(z) - i[\arg(\mp iz) - \arg(z) \pm\pi/2] \;.
\end{equation}

As $\ln^{(+)}(z)$ and $\ln^{(-)}(-z)$ are both defined on the complex plane cut along the negative imaginary axis they can differ only by a constant: $\ln^{(-)}(-z) = \ln^{(+)}(z) - i\pi$. In \eref{eq:expint} this gives
\begin{eqnarray}\label{eq:etaseries}
	\eta(z) & = & -[\gamma + \ln^{(+)}(z)]\sinh z - \frac{i\pi}{2} e^{-z} + \frac{1}{2} [ e^z \mathrm{Ein}(z) - e^{-z} \mathrm{Ein}(-z)] \\
	 & = & -[\gamma + \ln^{(+)}(z)]\sinh z - \frac{i\pi}{2} e^{-z} + z+\frac{11 z^3}{36}+\frac{137 z^5}{7200}+O\left(z^7\right) \nonumber \;,
\end{eqnarray}
from which can be seen that $\eta(z)$ still has a branch point at the origin, but, unlike $E_1(z)$, it remains finite there.

The following symmetry properties, reflection about the imaginary axis and reflection about the real axis (with exponential correction), can be proved from \eref{eq:etaseries} and may be used to continue $\eta ( z )$ out of any quadrant into the other three quadrants,
\begin{eqnarray}
	\eta ( z ) & = & - \eta( -z^* )^*  \;,\label{eq:ImAxisRefl}\\
	\eta ( z ) & = & \eta ( z^* )^* - i \pi e^{-z\,\sre z} \label{eq:ReAxisRefl} \;,
\end{eqnarray}
where
\begin{equation}\label{eq:sdef}
	\sre\,z \equiv \sgn(\Re\,z)
\end{equation}
and ${}^*$ denotes complex conjugation. These symmetries are apparent in the plots in \fref{fig:etaplots}.

The asymptotic expansion for large $|z|$ may be obtained from that for $f(z)$ \cite[\S 5.2.34]{Abramowitz_Stegun_72}
\begin{equation}\label{eq:etaasymp}
	\eta ( z ) \sim \frac{1}{z}\left(1+\frac{2!}{z^2}+\frac{4!}{z^4}+\ldots\right)
	\quad \mathrm{as}\quad |z| \to \infty, \quad -\frac{\pi}{2} < \Arg\, z < \frac{3\pi}{2} \;.
\end{equation}
However, the successive terms in this expansion do not decrease sufficiently rapidly for it to be very useful for computational purposes over the range of $|z|$ in which we are interested.

All the derivatives $\eta^{(n)} ( z)$ of $\eta ( z)$ may be expressed in terms of $\eta (z)$ and $\eta^{\prime} ( z )$,
\begin{eqnarray}\label{eq:etaderivs}
	\eta^{(n)}  (z) &=& \eta(z) - \frac{(n-2)!}{z^{n-1}} - \frac{(n-4)!}{z^{n-3}} -\ldots - \frac{0!}{z} \;;
	\quad n\,\mathrm{even} \;,\label{eq:etapreven}\\
	\eta^{(n)}  (z) &=& \eta^{\prime}(z) + \frac{(n-2)!}{z^{n-1}} + \frac{(n-4)!}{z^{n-3}} + \ldots + \frac{1!}{z^2}
	\;;\quad n\,\mathrm{odd} \label{eq:etaodd}\;.
\end{eqnarray}
The two relations above are used to calculate $\eta ( z)$ for intermediate $|z|$ by extrapolating from known values of $\eta ( z)$ and $\eta^{\prime}( z )$ using Simpson's rule. The known values are calculated from table 5.6 of reference \citen{Abramowitz_Stegun_72} using \eref{eq:expint} and \eref{eq:ReAxisRefl} and the relations:
\begin{eqnarray}
	\eta^{\prime}(z) &=& \frac{1}{2}[ e^z E_1^{(+)}(z) + e^{-z} E_1^{(-)}(z)]\\
	\eta^{\prime}(z) &=& \eta^{\prime}(z^*)^* + \sre(z) i \pi e^{-z\,\sre z} \;,
\end{eqnarray}
where $\sre(\cdot)$ is defined in \eref{eq:sdef} and
\begin{eqnarray}
	E_1^{(\pm)}(\pm z) &=& E_1 (\pm z) \quad\mathrm{for}\quad \Im\, z > 0 \;,\label{eq:E1pm}\\
	\quad\: E_1(z)^* &=& E_1 (z^*) \label{eq:E1star}\;.
\end{eqnarray}

For large $|z|$, Laguerre integration \cite[\S 25.4.45]{Abramowitz_Stegun_72} of \eref{eq:alt} is used,
\begin{equation}\label{eq:etaLaguerre}
	\eta (z) = z \sum_{i=1}^{n} \frac{w_{i}}{z^2 - x_{i}^{2} } + R_{n} ; \quad\Im\, z > 0 \;,
\end{equation}
\begin{equation}\label{eq:etaLagerr}
	|R_{n}| < \frac{(n!)^2 } { (\Im\, z)^{2n+1} } \;,
\end{equation}
with the abscissas $x_i$ and weight factors $w_i$ obtained from \cite[table 25.9]{Abramowitz_Stegun_72}. The error bound is pessimistic at large $|\Re\,z|$, as this equation provides accuracy of six significant figures, even for $z$ real, if $|\Re\,z| \ge 16$.

The FORTRAN IV subroutine developed for efficient calculation of $\eta(z)$ using the methods described in this appendix was listed in reference \citen{Dewar_67}. It is accurate to around 2 parts in $10^4$ and evaluates $\eta(z)$ in about $0.5\mu$s, so this innermost integral over $0 \leq k < \infty$ in the Fourier inversion may be regarded as performed analytically, leaving only the 2-dimensional integral over solid angle to be done numerically.

To produce the numerical results presented above we used this core subroutine in our reconstruction of the original screened field program (a testament to the backwards compatibility of FORTRAN). For the historical record we provide the FORTRAN source code, and a README file describing its use, as supplementary data in the online version of this paper. However, for accurate calculations it would probably be preferable to use the professionally written FORTRAN code for $E_1(z)$ developed by Amos \cite{Amos_90b}.

\setcounter{section}{1}

\section*{References}
\bibliographystyle{iopart-num}
\bibliography{DressedTestParticles}

\end{document}